\documentclass{iopart}

\usepackage{cite}

\def\vep{\varepsilon}

\def\al{\alpha}

\def\ep{\epsilon}
\def\be{\begin{equation}}
\def\ee{\end{equation}}
\def\bea{\begin{eqnarray}}
\def\eea{\end{eqnarray}}
\newcommand{\beqn}{\begin{eqnarray}}
\newcommand{\eeqn}{\end{eqnarray}}
\newcommand{\beqnn}{\begin{eqnarray*}}
\newcommand{\eeqnn}{\end{eqnarray*}}

\begin{document}

\title{Current status of the Dynamical Casimir Effect}

\author{V V Dodonov}

\address{
 Instituto de F\'{\i}sica, Universidade de Bras\'{\i}lia,
Caixa Postal 04455, 70910-900 Bras\'{\i}lia, DF, Brazil }

\eads{\mailto{vdodonov@fis.unb.br} }

\begin{abstract}

This is a brief review of different aspects of the so-called Dynamical Casimir Effect
and the proposals aimed at its possible experimental realizations. 
A rough classification of these proposals is given and important theoretical problems 
are pointed out.

\end{abstract}


\section{Introduction}

The term {\em Dynamical Casimir Effect\/} (DCE), 
introduced apparently
by Yablonovitch \cite{Yabl89}   and Schwinger \cite{Sch-DCE},
is frequently used nowadays
for the plethora of phenomena connected with the photon generation from vacuum due to 
fast changes of the geometry (in particular, the positions of some boundaries) 
or material properties of electrically neutral macroscopic
or mesoscopic objects
\footnote{Yablonovitch wrote: `\ldots we are considering
sudden nonadiabatic changes which have the effect of causing real transitions and boosting
the quantum fluctuations into real photons. In that sense this process may be called
the dynamic or nonadiabatic Casimir effect.'
Schwinger wrote: `\ldots I interpret as a dynamical 
Casimir effect wherein dielectric media are accelerated and emit light.'}. 
A rough qualitative explanation of such phenomena is the parametric amplification
of quantum fluctuations of the electromagnetic (EM) field in systems with time-dependent
parameters. The reference to vacuum fluctuations explains the appearance of Casimir's name 
(by analogy with the famous {\em static\/} Casimir effect, which is also considered frequently
as a manifestation of quantum vacuum fluctuations \cite{Milonni,Milton,Most09}), although
Casimir himself did not write anything on this subject. 
In view of many different manifestations of the DCE considered until now
it seems reasonable to make some rough classification. It is worth remembering that the
static Casimir effect has two main ingredients: quantum fluctuations and the presence of boundaries
confining the electromagnetic field. Therefore I shall use the
abbreviation MI-DCE ({\em Mirror Induced\/} Dynamical Casimir Effect) for those phenomena 
where the photons are created due to the movement of mirrors or changes of their material properties. 
An explicit connection between quantum fluctuations and the motion of boundaries
 was made in  \cite{Sarkar,DKM89},
where the name `Nonstationary Casimir Effect' was introduced,
and in \cite{BE,Lamb}, where the names `Mirror-Induced Radiation'  and `Motion-Induced Radiation' 
(with the same abbreviation MIR) were proposed.
The phenomena where the photons can be created due to the parametric amplification of vacuum fluctuations in media
without moving or changing boundaries will be referred to as PA-DCE
({\em Parametric\/} Dynamical Casimir Effect). 
Although these names contain an obvious tautology, I try to maintain the combination DCE because it
 became generally accepted by now.
I confine myself to the studies related to cavities, resonators or equivalent set-ups which could give a possibility
to verify the DCE experimentally, leaving aside many other special cases considered up to now, 
such as the single mirror examples or numerous situations where the amount of quanta created due to the DCE 
is obviously
too small to be detected. 
Extensive lists of publications on the DCE and systems with moving boundaries can be found in reviews \cite{D-rev1,DD-rev2,DCas60}.
Main theoretical ideas and predictions are briefly discussed in section \ref{sec-theor}.  
Concrete available experimental schemes are considered in section \ref{sec-exp}, where some
important problems waiting for the solution are also pointed out.

\section{Main ideas and theoretical predictions}
\label{sec-theor}

The most important requirements which must be fulfilled in order to observe the DCE
can be easily understood if one remembers the main idea laying in the foundation of the
theory of the electromagnetic field quantization, namely, that (roughly speaking) the EM field behaves
as a set of harmonic oscillators. Mathematically it is expressed by
writing the Hamiltonian operator of the free field in some cavity as
\begin{equation}
\hat {H}_0 = \sum_{n=1}^{\infty}\hbar\omega_n\left(\hat{a}^{\dag}_n\hat{a}_n+ 1/2\right)
\label{Ham}
\end{equation}
where $\hat {a}_n$ is the annihilation bosonic operator for the field mode
with frequency $\omega_n$. To change the energy of the field (in particular,
to create quanta from the initial vacuum state) at the level of formulas one must add to $\hat{H}_0$ some operator 
$\hat{H}_I(t)$ describing the interaction between the field and the `material' world.
Since the coupling between the field and `matter' is usually weak, $\hat{H}_I(t)$ can be
expanded in the series with respect to powers of the annihilation and creation operators.
The linear terms of this expansion describe the field excitation by external currents and
charges, which seems to be quite a classical effect. The second-order terms contain
combinations of the operators $\hat{a}^{\dag}_n\hat{a}_k$, $\hat{a}_n\hat{a}_k$ (and their
Hermitially conjugated counterparts) with time-dependent coefficients. These terms describe, in particular,
possible time-dependent changes of the eigenfrequencies and the squeezing effects \cite{Plun}.
 Physically, such interactions arise due
to changes of the cavity geometry or the electromagnetic properties of the medium filling in
the cavity. Such changes of parameters also result in the field excitation due to the amplification
of the initial fluctuations. In a sense, this is also a classical effect, except for the
important special case of the initial ground state: there are no fluctuations in this state
from the classical point of view, whereas `zero-point' fluctuations are predicted by quantum physics.
Therefore the photon creation from the initial vacuum state due to changes of parameters
is considered usually as a quantum effect whose experimental observation could be interpreted
as a `direct' proof of the existence of vacuum fluctuations (in contradistinction
to the `indirect' manifestations through the static Casimir effect, Lamb shift and many
other phenomena). This explains the attractiveness of the DCE for many researchers, both
theoreticians and experimentalists.

Following this line of reasoning one may suppose that significant features of the
phenomenon could be caught in the simplest example of a {\em single\/} harmonic quantum oscillator
(representing some field mode) with a time-dependent frequency \cite{Man91}.
The theory of quantum nonstationary harmonic oscillator has been well developed since its foundation
by Husimi in 1953 \cite{Hus} (see, e.g., \cite{book} for the review and references).
It appears that all dynamical properties of the {\em quantum\/} oscillator are determined
by the fundamental set of solutions of the {\em classical\/} equation of motion
\be
\ddot{\vep} +\omega^2(t) \vep=0.
\label{claseq}
\ee
In particular, if $\omega(t) = \omega_i $ for $t<0$
and $\omega(t) = \omega_f $ for $t>t_f$,
then the information
on the state of the quantum oscillator at $t>t_f$
is encrypted in the complex coefficients
$\rho_{\pm}$
of the asymptotic form of the solution
$\vep({t>t_f}) =\omega_f^{-1/2}\left[\rho_{-} e^{-i\omega_f t} +
\rho_{+} e^{i\omega_f t}\right]$
originated from
$\vep({t<0}) = \omega_i^{-1/2} e^{-i\omega_i t}$.
For the initial thermal state of
temperature $\Theta$ the mean number of quanta at $t>t_f$ 
 equals \cite{Hus,DKNPR,Plun00}
\be
{\cal N} \equiv \langle \hat{a}^{\dag}\hat{a}\rangle
=G\left(\frac{|\dot\vep|^2 + \omega_f^2|\vep|^2}{4\omega_f} - \frac12\right)
= G \frac{R}{T}
\label{N-vep}
\ee
where $G= \coth\left[{\hbar\omega_i}/{(2k_B \Theta) }\right]$. The quantities
$R\equiv |{\rho_{+}}/{\rho_{-}}|^2$ and $T\equiv 1-R\equiv |\rho_{-}|^{-2}$
can be interpreted as energy reflection and transmission
coefficients from an effective `potential barrier'
given by the function $\omega^2(t)$.

The intuition suggests that for the {\em monotonic\/} function $\omega(t)$ 
the effective reflection coefficient cannot exceed the value given by the Fresnel
formula for the instantaneous jump of the frequency: 
$R\le \left(\omega_i - \omega_f\right)^2/\left(\omega_i + \omega_f\right)^2$. 
This suggestion was proved rigorously in \cite{Viss}.
If the frequency varies due to the change of the cavity characteristic dimension $L$,
then for small variation $\Delta L$ one obtains an estimation of the maximal
possible number of created photons in a single mode ${\cal N}_{max} \sim (\Delta L/L)^2$.
Besides, the variations must be fast, because the number of quanta is the adiabatic
invariant, so the photons cannot be created in slow processes. The duration of the fast
process must be of the order of the period of the field oscillations (faster motions
do not result in a significant increase of the number of photons, which is limited, after all,
by the total change of frequency). This gives another estimation
${\cal N}_{max} \sim (v/c)^2$, where $v$ is the characteristic velocity of the 
boundary and $c$ the speed of light. Consequently, the DCE is the relativistic effect
of the second order and the expected number of created photons in a single mode is much less
than unity for monotonic nonrelativistic motions. 

This fact shuts down a possibility that the phenomenon of {\em sonoluminescence \/}
(emission of bright short pulses of the visible light from air bubbles in the water 
when the bubbles pulsate due to the pressure oscillations in a strong standing acoustic wave
 \cite{sono-Barb,sono-Bren}) could be related to the DCE, although it was
the starting point of Schwinger's research \cite{Sch-DCE}.
Real hydrodynamical processes in the bubbles are too slow (even at the picosecond time scale)
compared with fast oscillations of electromagnetic fields at the optical frequencies,
so that the motion of bubble's surface should be considered as adiabatic from the
electrodynamic point of view. Since the effective reflection coefficient is much smaller
than unity for adiabatic processes \cite{DK96}, the mean number of photons created due to the bubble
pulsations must be many orders of magnitude less than unity.
(Besides, the actual change of the `vacuum energy' due to the variations of the bubble size turns out
to be ten orders of magnitude smaller than the initial Schwinger's evaluations \cite{Milt04}.)

However, it is well known that the effective reflection coefficient can be made as close to
unity as desired in the case of {\em periodic\/} variations of parameters satisfying some
{\em resonance conditions\/} (by analogy with periodic spatial structures). 
This is the basis of the proposals 
to use the {\em parametric amplification\/} effect in experiments on the DCE \cite{Lamb,Man91,Plun00,DK96,DK92}.
Earlier, this idea was put forward in \cite{Sarkar,Riv79}, but the evaluations of the
effect were not correct.
A possibility of a significant amplification of the {\em Casimir force\/}  under
the resonance conditions was pointed out in \cite{Brag} for the $LC$-contour 
and in \cite{Jaekel} for the Fabry--P\'erot cavity.

For harmonic variations of the frequency in the form
$\omega (t)=\omega_0\left[1+2\kappa\cos(2\omega_0t)\right]$ with $|\kappa| \ll 1$
equation (\ref{claseq}) can be solved approximately using, e.g., the method of averaging over fast oscillations
or the method of slowly varying amplitudes
\cite{LL-book}. Then equation (\ref{N-vep}) yields \cite{DKNPR,DK96}
\begin{equation}
{\cal N} =\sinh^2\left(\omega_0\kappa t\right).
\label{num-1}
\end{equation}
This formula does not take into account inevitable losses due to the dissipation in the cavity walls.
For the cavity with a finite quality factor $Q=\omega_0/(2\gamma)$ (where $\gamma$ is the {\em amplitude\/}
damping coefficient) one could suppose that formula (\ref{num-1}) can be
reliable for $\gamma t\ll 1$, so that the maximal number of photons could be roughly evaluated by putting
$t=1/\gamma$ in (\ref{num-1}):
${\cal N}_{max} \sim\sinh^2\left(2Q\kappa \right)$, meaning that the necessary condition for the
creation of more than one photon is $2Q\kappa >1$. The second statement is correct while the estimation of
${\cal N}_{max}$ is not. 
Indeed, the calculations made in the framework of the linear master equation with the standard dissipative
superoperator \cite{SaHy96b,D98} gave the following asymptotical formula for the number of photons created from the
initial thermal state for $2\omega_0\kappa\zeta t >1$ \cite{D98}:
${\cal N} \approx (4\zeta)^{-1}\exp(2\omega_0\kappa\zeta t)$, where $\zeta=1-(2Q\kappa)^{-1}$. 
Consequently, the exponential growth of the number of photons is possible if $2Q\kappa >1$. 
Of course, such a fast growth cannot continue forever, since the linear approximation becomes invalid
if $\omega_0\kappa^2 t \sim 1$. First estimations of the saturated value of the photon number due to
nonlinear effects were made in \cite{Sriv06}, but
this problem needs further investigations.

However, at the current moment the paramount task is to observe at least the beginning of the process
of the photon generation. Although formula (\ref{num-1}) provides some insights, the real process
is more complicated, in particular, due to the presence of infinitely many field modes in the cavity. 
How this circumstance could influence the rate of the
photon production in each mode and the total number of created quanta? 
A powerful tool for answering this question (although partially, neglecting the dissipation)
is the method of {\em effective Hamiltonians\/} \cite{Plun,Law94}. It can be formulated as follows. 
 Suppose that the set of Maxwell's
equation in a medium with {\em time-independent\/} parameters and
boundaries can be reduced to an equation of the form
$\hat{\cal K}(\{L\}){\bf F}_{\al}({\bf r};\{L\})=
\omega_{\al}^2(\{L\}) {\bf F}_{\al}({\bf r};\{L\})$,
where $\{L\}$ means a set of parameters 
(for example, the distance between the walls or the dielectric permittivity inside the cavity),
 $\omega_{\al}(\{L\})$ is the eigenfrequency
of the field mode labeled by the number (or a set of numbers) $\alpha$
and ${\bf F}_{\al}({\bf r};\{L\})$ is some vector function describing
the EM field (e.g., the vector potential). 
 In the simplest cases $\hat{\cal K}(\{L\})$ is reduced to
the Laplace operator.
Usually, the operator $\hat{\cal K}(\{L\})$ is self-adjoint, and the set
of functions $\{{\bf F}_{\al}({\bf r};\{L\})\}$ is orthonormal and
complete in some sense.

Now suppose that parameters $L_1,L_2,\ldots,L_n$ become time-dependent.
If one can still satisfy automatically the boundary conditions,
expanding the field ${\bf F}({\bf r},t)$ over `instantaneous'
eigenfunctions
$
{\bf F}({\bf r},t)=\sum_{\al} q_{\al}(t) {\bf F}_{\al}({\bf r};\{L(t)\})$
(this is true, e.g., for the Dirichlet boundary conditions,
which are equivalent
in some cases to the TE polarization of the field modes),
then the dynamics of the field is described completely by the generalized
coordinates $q_{\al}(t)$, whose equations of motion
can be derived from the
{\em effective time-dependent Hamiltonian\/} \cite{Plun}
\be
H = \frac12\sum_{\al}\left[p_{\al}^2
+ \omega_{\al}^2\left(\{L(t)\}\right) q_{\al}^2\right]
+ \sum_{k=1}^n \frac{\dot{L}_k(t)}{L_k(t)}\sum_{\al \neq \beta} p_{\al} m_{\al\beta}^{(k)}q_{\beta},
\label{genHam}
\ee
\be
m_{\al\beta}^{(k)} = - m_{\beta\al}^{(k)} = L_k \int dV
\frac{\partial {\bf F}_{\al}\left({\bf r};\{L\}\right)}{\partial L_k}
{\bf F}_{\beta}\left({\bf r};\{L\}\right).
\label{mab}
\ee
Consequently, the field problem can be reduced to studying the dynamics of the infinite set of harmonic
oscillators with time-dependent frequencies and bilinear specific (coordinate--momentum)
time-dependent coupling. The preceding one-mode example shows that the most important (from the point of view
of applications to the DCE) cases are those where the parameters $L_k(t)$ vary in time {\em periodically\/}.
In the case of small {\em harmonic\/} variations at the frequency close to the double unperturbed eigenfrequency
of some mode $2\omega_0$, the equations of motion resulting from Hamiltonian (\ref{genHam}) can be solved approximately
with the aid of the method of slowly varying amplitudes. If the difference $\omega_{\alpha} -\omega_{\beta}$ is not close
to $2\omega_0$ for all those modes which have nonzero (or not very small) coupling coefficients $m_{\al\beta}$,
then only the selected mode with label $0$ can be excited {\em in the long-time limit\/}, and one can
consider only {\em single resonance mode\/} \cite{DK96}. However, the intermode coupling can be important
in some cases, especially for large amplitudes of the frequency variation.

For example, the {\em resonance coupling\/} between two modes is possible 
in cubical cavities \cite{Croc1}. This case was studied in \cite{Croc1,AVD}.
It was shown that the number of photons in both the coupled modes grows exponentially with time
in the `long time' limit $\omega_0\kappa t \gg 1$, but the rate of photon generation (the argument
of the exponential function) turns out to be {\em twice smaller\/} than the value of this rate
 in the absence of the resonance coupling. (Actually, this rate depends on the concrete values of the
coupling coefficients $m_{\al\beta}$, but in any case it cannot exceed the `uncoupled' values \cite{AVD}.)
 This example indicates that the resonance coupling
between the modes should be avoided in order to achieve the maximal photon generation rate, at least
in the case of TE modes. A detailed numerical study of this case for different sizes of the
rectangular cavities was performed in \cite{Ruser06-PRA}.
 The authors of the recent paper \cite{Naylor09} 
used numerical methods taking into account the interaction
between 50 lowest coupled modes in the rectangular cavity bisected by a `plasma sheet'
with a periodically varying number of free carriers. Some plots in that paper show that the intermode coupling
can increase the number of photons in the modes of EM field with the TM polarization. But the maximal
number of created photons in that examples did not exceed $5$ (in contrast with \cite{Ruser06-PRA} where
the limits of time integration were extended to much bigger values, permitting to reach the regime of
large numbers of created photons). Therefore more precise calculations
for a larger time scale are necessary in the TM case and for other geometries.

In the most distinct form the `destructive' role of coupling between the modes can be seen in the example of
a one-dimensional cavity with an ideal moving boundary. The first calculations of the number of created
photons in this case were made by Moore \cite{Moore} 40 years ago, although the dynamics of classical 
EM fields in this geometry was studied by many authors since 1920s  \cite{Hav,Nic2,Bar67,Ves-app}.
One of possible physical realizations of this model is the Fabry--P\'erot resonator;
another possibility is the TEM modes in a coaxial cylindrical cavity \cite{Croc05}.
The specific feature of this model is the {\em equidistant} form of the spectrum
of eigenfrequencies: $\omega_n=c\pi n/L$.
Namely for this reason
the Heisenberg equations of motion following from the Hamiltonian (\ref{genHam}) can be reduced
to a simple set of equations admitting analytical solutions \cite{DK96,Djpa}.
Numerical calculations made in \cite{Ruser05,Ruser06} confirmed
a high accuracy of these analytical solutions.
It was shown in \cite{DK96,Djpa} that 
the number of photons created from vacuum in the $n$th (odd) mode ${\cal N}_n$ 
depends on time $t$ {\em linearly\/} in the asymptotical regime $\kappa\omega_1 t \gg 1$, 
whereas the total number of photons in all the modes 
${\cal N}_{tot} = \sum_{n=1}^{\infty}{\cal N}_n$ grows with time quadratically:
\be
{\cal N}_n \approx 8\kappa\omega_1 t/(\pi^2 n), \quad 
{\cal N}_{tot}  \approx 2(\kappa\omega_1 t)^2.
\label{Nlin}
\ee
The total energy 
${\cal E} = \sum_{n=1}^{\infty} \hbar \omega_n {\cal N}_n$
increases exponentially, 
${\cal E} =(\hbar\omega_1/4)\sinh^2(2\kappa\omega_1 t)$, due to the exponential
increase of the number of excited modes 
\cite{Djpa,Klim97}. 

On the other hand, just due to the mode coupling some interesting phenomena could be observed
in cavities with equidistant spectra, such as, for example, the formation of narrow packets, both
inside the cavity (where they bounce periodically  between the walls \cite{Ves-osc,LawPRL,Cole,Dal99,AD00}) 
and outside it \cite{Lamb-puls,Lamb98}.
For this reason attempts to observe the DCE in such cavities are quite interesting, too.

\section{Experimental proposals for observing the DCE}
\label{sec-exp}

\subsection{Difficulties with real moving boundaries}

According to formula (\ref{num-1}) (confirmed
by several groups using different analytical \cite{Plun00,Croc1,Croc05} 
and numerical \cite{Ruser06-PRA} approaches),
a possibility of experimental verification of the DCE depends on the 
amplitude of the frequency variation $\Delta\omega= 2\kappa\omega_0$.
The main difficulty is due to the very high frequency
$2\omega_0$. The most exciting dream is to observe the `Casimir light' \cite{Sch-DCE}
in the visible part of the electromagnetic spectrum. But it seems to be very improbable,
at least in the `pure' form, using real mirrors oscillating at the frequency about $10^{15}\,$Hz.
Indeed, let us consider a suspended metallic plate of density $\rho$, area $S$ and thickness $b$, illuminated
by the laser beam of frequency $\omega_0$ and average intensity $I$. The radiation pressure force 
depends on time as $F(t) = 2IS\left[1 +\cos(2\omega_0 t)\right]/c$ (for the uniform illumination), 
so it can cause the forced oscillations of the plate exactly at twice the frequency $2\omega_0$.
Since the optical frequency $\omega_0$ is many orders of magnitude higher than the mechanical
frequency of the suspension, the amplitude of displacements of the plate from the mean position equals
$\Delta L = F_{max}^{(oscil)}/[m(2\omega_0)^2] = I/(2c\rho b \omega_0^2)$. It results in the 
amplitude of the frequency variation 
$\Delta\omega=\xi \omega_0(\Delta L/L)= \xi(\lambda/L)I/(4\pi c^2 \rho b)$,
where $\lambda=2\pi c/\omega_0$ is the wavelength in vacuum corresponding to frequency $\omega_0$,
$L$ is the average value of the variable length of the cavity and $\xi$ is the numerical factor
of the order of unity ($\xi=1$ for the Fabry-P\'erot cavity modeled by two infinite parallel plates).  
Taking $\lambda/L \sim 1$, 
$\rho \sim 3\times10^3\,$kg$\,$m$^{-3}$ (Al) and $b\sim 1\,\mu$m one can see that the 
frequency variation amplitude $\Delta\omega=1\,$s$^{-1}$ can be achieved for the laser intensity
$I\sim 3\times10^{15}\,$W$\,$m$^{-2}$. According to formula (\ref{num-1})
the product $\Delta\omega t$ should be not smaller than unity to generate more than one photon in the mode.
Consequently the total laser energy per unit area of the plate
should be not less than $3\times10^{15}\,$J$\,$m$^{-2}$, which is obviously unrealistic.
This estimation shows the
impossibility of exciting and maintaining the {\em high frequency oscillations\/} of the suspended plate with
a large amplitude and for a long time. On the contrary, the {\em low frequency\/}
oscillations at the mechanical frequency of the suspension can be excited. They result,
in particular, in the Kerr-like back-action effect on the field 
\cite{Fabre94}. This is a very interesting area, including 
the generation of the so-called `nonclassical states' (e.g., quantum superpositions)
of the field and the mirror (considered as a quantum object) \cite{Manci,Bose,Marsh03},
the mirror--field entanglement \cite{Manci02,Zhang03,Esli06,Ferr06},
cooling mirrors by the radiation pressure \cite{Coha99,Gig06,Arc06,Kleck06,Huang09}, 
etc., but it is totally distinct from the DCE.

It seems that the only possible way to realize the real
motion of material boundaries at high frequencies is not to Q2
move the whole mirror, but to cause its surface to perform
harmonic vibrations with the aid of some mechanism, e.g.
using the piezo-effect \cite{D95}.
The amplitude of such vibrations $\Delta L$ is connected with the maximal relative deformation
$\delta$ in a standing acoustic wave inside the wall as
 $\delta =\omega_w \Delta L /v_s$, where $v_s\sim 5\cdot 10^3$~m/s is the sound velocity.
 Since usual materials cannot
bear deformations exceeding the value $\delta_{max}\sim 10^{-2}$,
the velocity of the boundary cannot exceed the value 
$v_{max}\sim \delta_{max}v_s \sim 50$~m/s (independent on the frequency). The maximal possible
frequency variation amplitude $\Delta\omega$ can be evaluated as
$\Delta \omega = \xi v_s \delta_{max}/(2L)$. For the optical frequencies $2L >1\,\mu$m and
$\Delta \omega < 5\times 10^7 \,$s$^{-1}$, whereas for the microwave frequencies (in the GHz band)
$L \sim 1\,$cm and $\Delta \omega <  10^3 \,$s$^{-1}$. Since the time of excitation $t$ must be bigger
than $1/\Delta\omega$, the quality factor of the cavity $Q$ must be not less than 
$Q_{min} \approx \omega_0/\Delta\omega \approx (L/\lambda)4\pi c/(v_s \delta) \sim 10^8 (L/\lambda)$.
Consequently, there are two main challenges: how to excite high frequency surface oscillations
and how to maintain the high quality factor in the regime of strong surface vibrations.
The excitation of high amplitude surface vibrations at the optical frequencies seems very problematic.
Therefore hardly the `Casimir light' in the visible region can be generated in systems with really
moving boundaries. However, this seems to be possible in other schemes, where changes of some parameters
can be interpreted as variations of an `effective length' of the cavity: see subsection \ref{sec-out}. 

The GHz frequency band seems more promising. In such a case the
dimensions of cavities must be of the order of few centimeters.
Superconducting cavities with the quality factors exceeding $10^{10}$ in the frequency band 
from $1$ to $50$ GHz are available for a long time \cite{Wal90,Padam01,Kuhr07}. 
Therefore the most difficult problem is to excite the surface vibrations. At lower frequencies
it was solved long ago. For example, the excitation of vibrations of the mirror 
at the frequency $60\,$MHz with the aid of a quartz transducer was reported 
in \cite{Henneb}. The calculated values of the peak displacement and velocity were $1.4\times 10^{-8}\,$cm
and $5.3\,$cm/sec. Recently a significant progress was achieved in fabrication of the so called
`film bulk acoustic resonators' (FBARs): piezoelectric devices working at the frequencies from
1 to 3 GHz \cite{Tadig09}. They consist of an aluminum nitride (AlN) film of thickness 
corresponding to one half of the acoustic wavelength, sandwiched between two electrodes.
It was suggested  \cite{Onof06} to use such kind of devices to excite the surface vibrations of cavities
in order to observe the DCE. However, the problem is very difficult,
and no experimental results in this direction were reported by now.

\subsection{Effective moving boundaries: MIR experiment with semiconductor mirrors}
\label{sec-MIR}

In view of difficulties of the excitation of oscillating motion of real boundaries,
the ideas concerning the {\em imitation\/} of this motion attracted more and more attention with the course
of time. 
The first concrete suggestion was made  two decades ago by Yablonovitch
\cite{Yabl89}, who proposed to use a medium with a rapidly
 decreasing in time refractive index (`plasma window')
 to simulate the so-called  Unruh effect.
Also, he pointed out that fast changes of dielectric properties can be
achieved in semiconductors illuminated by subpicosecond optical pulses
and supposed that `the moving plasma front can act as a moving mirror
exceeding the speed of light.'
Similar ideas and different possible schemes based on fast changes of the carrier concentration 
in semiconductors illuminated by laser pulses were discussed in 
\cite{Yabl89-2,Oku95,Loz}.
Yablonovitch \cite{Yabl89,Yabl89-2} put emphasis on the excitation of {\em virtual electron--hole pairs\/} by 
 optical radiation tuned to the transparent region just below the band gap
in a semiconductor photodiode. He showed that big changes of the {\em real part\/} of the dielectric
permittivity could be achieved in this way.

The key idea of the experiment named `MIR', which is under preparation in the university of Padua
 \cite{Padua,Padua05}, is to imitate the motion of a boundary, using an effective `plasma mirror'
formed by {\em real electron--hole pairs\/} in a thin film near the surface of a semiconductor slab,
illuminated by a periodical sequence of short laser pulses.
If the interval between pulses exceeds the recombination time of
carriers in the semiconductor, a highly conducting layer will
periodically appear and disappear on the surface of the slab.
This can be interpreted as periodical displacements of the boundary.
The basic physical idea was nicely explained in \cite{Padua05}: `\ldots this effective motion is
much more convenient than a mechanical motion, since in a metal mirror only the conduction electrons 
reflect the electromagnetic waves, whereas a great amount of power would be
wasted in the acceleration of the much heavier nuclei.'

The main advantage of the semiconductor mirror is a great increase of the maximal frequency shift,
compared with the case of vibrating surface. This shift is determined mainly by the thickness of the
semiconductor slab. Using the slabs of few millimeters thickness one can easily obtain the frequency
 variation amplitude $\Delta\omega \sim 10^7\,$s$^{-1}$ or even bigger in the GHz range of the
cavity resonance frequencies. Then the total excitation time can be reduced to less than $1\,\mu$s
and the cavity quality factor can be lowered to the easily achievable values of the order of $10^5$ 
or even $10^4$.
There are proposals \cite{Naylor09} to put the semiconductor slab in the middle of the
rectangular cavity. In this case the frequency shift attains the maximal value.
However, it is not quite clear whether the strong intermode coupling will
not diminish the final number of photons. Besides, in this configuration one can meet
problems with the uniform illumination of the slab and especially with heat removing.
Another possibility to increase the amplitude of the frequency shift variation and the photon
production rate is
to excite the TM cavity mode instead of the TE one \cite{Naylor09,Mund,Croc2,Plun04,DD-job05,DDjpa06}.
However, there is one technical difficulty which can be decisive from the experimental
point of view: in the TE case one can use 
the lowest TE$_{101}$ mode, whereas the TM$_{101}$ mode does not exist (the lowest one is TM$_{111}$). 
For this reason the
minimal area to be illuminated in the TM case is much bigger than in the TE configuration. In the example
considered in \cite{Naylor09} one has to illuminate uniformly very big cross section $5\times5\,$cm
from the distance $5\,$cm, whereas using the optimized rectangular geometry for the TE mode in 
the MIR experiment one can reduce the illuminated area to few cm$^2$, and even smaller area 
is necessary in cavities having more elaborate shapes.

The thickness of the photo-excited conducting layer nearby the surface of the semiconductor slab 
is determined mainly by the absorption coefficient of the laser radiation,
so it is about few micrometers or less (depending on the laser wavelength), being much smaller
than the thickness of the slab itself. Therefore laser pulses with the surface
energy density about few $\mu$J/cm$^{2}$ can create a highly conducting layer with the carrier concentration
exceeding $10^{17}\,$cm$^{-3}$, which gives rise to an almost maximal possible change of the cavity eigenfrequency 
for the given geometry \cite{Padua,DD-JPB}.
It is worth noting that although the thickness of the conducting layer is less than the skin depth, it gives the
same frequency shift as the conductor filling in all the slab. This interesting fact was explained and
verified experimentally in \cite{Pad-freq}.

Note also that the laser wavelength $\lambda_{las}$ is of the order of
$1\,\mu$m, while the wavelength $\lambda_{cav}$ of the fundamental
cavity mode which is supposed to be excited due to the DCE is about $10\,$cm.
Consequently, if an antenna put somewhere inside the cavity
and tuned to the resonance cavity frequency will register
a strong signal after the set of laser pulses, one can be sure that
the quanta of EM field in the fundamental mode were created due
to the DCE and they do not belong to some `tail' of the laser
pulse, just due to the difference by five orders of magnitude
between $\lambda_{las}$ and $\lambda_{cav}$.

However, using the semiconductor mirror in the DCE experiments one has to overcome several serious difficulties,
resulting from the fact that laser pulses create pairs of {\em real\/} carriers which
change mainly the {\em imaginary part\/} $\ep_2 \equiv {4\pi  \sigma(\omega)}/\omega$  
of the complex dielectric permittivity 
$\ep =  \ep_1 +i\ep_2$. Here $\sigma(\omega)$ is the real conductivity at frequency $\omega$
(in the CGS system of units). 
For example, let us use the simple Drude model formula
$
\ep(\omega) = \ep_a + {4\pi i \sigma_0}/{[\omega(1-i\omega\tau)]}$,
where a real constant $\ep_a$ describes the contribution of bounded electrons and ions,
$\sigma_0 = n e^2 \tau/m$  is the static (zero-frequency) conductivity,
$n$ is the concentration of free carriers (created by laser pulses) with charge $e$ and
 effective mass $m$ and $\tau$ the relaxation time.
The imaginary part of $\ep$ can be neglected under the condition $\omega\tau \gg 1$,
which means that the low-frequency mobility $b=|e|\tau/m$ (related to the low-frequency
conductivity $\sigma_0$ as $\sigma_0=n|e|b$) must be much bigger than $b_*(\omega)=|e|/(m\omega)$.
For the optical frequencies $\omega \sim 3 \times 10^{15}\,$s$^{-1}$ and for $m\sim m_e$ (the mass of free
electron) one has $b_*(\omega) \sim 5\times 10^{-5}\,$m$^2$V$^{-1}$s$^{-1}$, so that the condition
$b \gg b_*$ can be easily fulfilled, meaning that one can use the {\em real-valued\/} function
$\ep(\omega)$.
Namely this special case was considered by several authors
\cite{Plun04,Croc04,Naylor09} who studied quantum effects caused by the periodical variations
of properties of thin {\em ideal\/} dielectric slabs 
or infinitely thin {\em ideal\/} conducting
films (described by means of time-dependent $\delta$-potentials in the framework of the `plasma sheet' model)
put inside the resonance cavities. 
Unfortunately, the results of those studies, being interesting by themselves, cannot be applied
to the MIR experiment, where the resonance frequency is about $2.3\,$GHz 
($\omega \approx 1.4\times 10^{10}\,$s$^{-1}$). For this
frequency one obtains $b_*(\omega) \sim 10\,$m$^2$V$^{-1}$s$^{-1}$, whereas the reported values
of the mobility in the highly doped GaAs samples used in this experiment are of the order of
$1\,$m$^2$V$^{-1}$s$^{-1}$ \cite{Pad-rec,Pad-Cas60}, and hardly this mobility can be increased by two orders of magnitude
(maintaining the necessary very small recombination time)
to satisfy the condition $b \gg b_*$.
Therefore a much more reliable approximation of the complex dielectric function $\ep(\omega)$ 
which should be used in the analysis of realistic DCE experiments with
semiconductor time-dependent mirrors is
$\ep(\omega) = \ep_a + 4\pi i \sigma_0/\omega$.
\footnote{Note however that the results of \cite{Plun04} could be useful in experiments based on the effect
of {\em virtual photoconductivity\/}  \cite{Yabl89-2}, because $\ep(\omega)$ remains real in such a case. 
But it is unclear whether the `plasma sheet' model used in \cite{Croc04,Naylor09} can be applied to this case,
because this model was justified in \cite{Naylor09,BarCal95} for real free carriers.}
As a consequence, the `instantaneous' time-dependent resonance frequency becomes complex-valued function
$\Omega =\omega -i\gamma$.
The calculations made in \cite{DD-rev2,DCas60,DD-JPB} gave the following formulas for the time-dependent damping
coefficient $\gamma(t)$ and the frequency variation $\chi(t)=\omega(t)-\omega_0$:
\be
\chi(t) =\frac{\chi_{m} A^2(t)}{A^2(t)+1}, \quad 
\gamma(t) =\frac{|\chi_{m}| A(t)}{A^2(t)+1},
\label{chi-A}
\ee
where $\chi_{m}$ is the maximal possible frequency shift attained when the slab becomes a perfect conductor.
The function $A(t)$ is proportional to  the time-dependent integral of the free carriers concentration
across the slab. 
For ultrashort pulses and negligible surface recombination and diffusion coefficients,
 $A(t)=A_0\exp(-t/T_r)$
where $T_r$ is the recombination time and $A_0$ is proportional to the product of the
total energy of the laser pulse by the mobility of carriers.
Equation (\ref{chi-A}) clearly shows that although the damping coefficient $\gamma(t)$ 
can be safely neglected if $\ep_2 \ll 1$ and $A\ll 1$ (an almost ideal dielectric)
or $\ep_2 \gg 1$ and $A\gg 1$ (an almost ideal conductor), it becomes very important in the
intermediate regime, when the high concentration of carriers achieved after the action of a short
laser pulse returns continuously to the initial value. Even if $A_0\gg1$, during some time interval one has
$A(t) \sim 1$ and $\gamma(t)=|\chi_{m}|/2 \approx |\chi(t)|$.
Therefore the influence of dissipation is predominant at the final stages of the recombination process.
These observations show that without taking into account inevitable losses inside
the semiconductor slab during the excitation-recombination process
one cannot predict the results of the realistic DCE experiments for microwaves even qualitatively.

A simple model taking into account the dissipation was developed in \cite{DD-rev2,DCas60,DD-JPB,D09}.
It was assumed that the dynamics of a single nonstationary quantum
oscillator (representing the resonance mode of the field) with a time-dependent linear damping 
can be described in the frameworks of the Heisenberg--Langevin equations 
with two non-commuting and delta-correlated time-dependent noise operators. 
One of results is the following formula for the maximal mean number of photons
which could be generated from the initial
thermal field state after $n \gg 1$ pulses of periodicity $T$: 
\be
{\cal N}_{n} \approx  
\frac{G_f(\nu - \Lambda) + G_w\Lambda}{4(\nu - \Lambda)}
\, e^{2n(\nu -\Lambda)}.
\label{Ntotas}
\ee
Here
$\nu \approx  \left| \int_0^T  {\chi}(t) \exp(-2i\omega_0 t) dt\right|$
and $\Lambda = \int_0^T  {\gamma}(t)  dt$.
Note that the
initial temperatures and corresponding amplification coefficients of the field mode $G_f$
and the cavity walls $G_w$ can be different.
The exact value of $T$ must be close to the half-period of the
excited field mode but not coincide exactly with this half-period, in order to fulfil the resonance conditions.
 Numerical calculations show that the difference $\nu-\Lambda$
can be positive if only the energy of laser pulses exceeds some critical value \cite{DD-rev2,DD-JPB,DCas60}.  
 The existence of this critical value takes its origin in the
different behaviors of the real and imaginary parts of the frequency shift in the
semiconductor with {\em real\/} free carriers: for small concentrations of created carriers
(i.e., for low pulse energies) the imaginary part increases linearly as function of energy,
whereas the real part increases quadratically, as can be seen, in particular, in equation (\ref{chi-A}).
Numerically this critical value turns out rather high: different estimations give the values from $1\,\mu$J
to $10\,\mu$J or even $100\,\mu$J, depending on the cavity geometry, recombination time and mobility
of carriers. But the decisive factor is the energy gap of the semiconductor of an order of $1\,$eV.

For $A_0 \gg 1$ one can obtain \cite{D09} simple approximate formulas
$\Lambda \approx \pi |\chi_{m}|\omega_0T_r /2$ and
\be
2\nu/|\chi_m| \approx \left| 1
-\frac{\pi \omega_0 T_r \exp\left(-2i\omega_0 T_r \ln A_0 \right)}{\sinh(\pi\omega_0 T_r)}
\right|.
\ee
They show that the photon generation can be achieved only for short recombination
times $T_r<0.5\omega_0^{-1}$. This requires a hard work on preparing the semiconductor samples
satisfying contradictory requirements: a short recombination time (less than
 $20$ ps)  but a high mobility.
Nonetheless it seems that these problems can be resolved 
\cite{Pad-rec,Pad-Cas60} and the first experiment on the MI-DCE will be done soon.
It is expected that 1000--2000 laser pulses will be sufficient to generate several thousand microwave photons,
much more than the measured sensitivity level of about 100 photons \cite{Brag09}.

\subsection{MI-DCE with illuminated superconducting boundaries}

Some of problems mentioned in the preceding subsection can be softened if laser pulses illuminate not the semiconductor
but {\em superconductor\/} surfaces. In this case the changes of dielectric properties happen due to the
transition from the superconducting to normal conducting phase caused by the local heating of the surface.
Since the energy gap in superconductors is several orders of magnitude smaller than the energy gap
in semiconductors, the energy of laser pulses can be made several orders of magnitude smaller than
in the case of semiconductor mirrors. 
The frequency modulation of the superconducting  microwave resonator by laser irradiation
was reported in \cite{Tsin94}. The authors of that paper used
the parallel-plate resonator consisting of two superconducting YBa$_2$Cu$_3$O$_{7- x}$  films of
$300\,$nm thickness with dimensions $10\times 10\,$mm$^2$,
 separated by a sapphire spacer of the thickness $0.3\,$mm. The temperature varied between $30\,$K up 
to $T_c \approx 91\,$K. They illuminated one plate by the cw Ar laser 
with the diameter of illuminated spot about $1\,$mm and demonstrated the downwards shift of the cavity
 resonance curve by $0.5\,$MHz from the initial value $5400\,$MHz 
 without a visible change of the
width of the curve. This suggests that the imaginary part of the resonance frequency change can be
neglected. 
Therefore the schemes based on changing electrodynamic properties of superconducting mirrors seem
to be rather promising.

\subsection{Various PA-DCE schemes}

A concrete proposal relating DCE with the laser illuminated superconductors was made in \cite{Seg07},
where the superconducting {\em stripline resonator\/} (a ring having a radius of 6.39 mm 
and width $347\,\mu$m composed of NbN film of $8\,$nm thickness deposited on a Sapphire wafer 
\cite{Arbel06}) was considered as a promising candidate for the photon generation from vacuum
in the range from 2 to 8 GHz.
Strictly speaking, it is difficult to connect the change of the frequency shift of this resonator
with an effective motion of some boundary. But if one assumes the definition of the DCE as the
phenomenon of photon creation from vacuum due to the change of {\em some\/} parameters of a system \cite{Man91},
then this scheme fits perfectly to the PA-DCE family.
The advantages of proposals based on the periodic illumination of superconductors consist in
the easy modulation of the resonance frequency,
the big amplitude of its variations and a low necessary energy of laser pulses.
For example, a parabolic dependence of the frequency shift on the pulse energy was reported in 
\cite{Cho03}. The $70\,$ps pulses of the energy $3\,$nJ resulted in the $20\,$MHz shift at the temperature
$20\,$K and almost $100\,$MHz at $80\,$K (for the YBa$_2$Cu$_3$O$_{7- x}$ strips).
The NbN films demonstrated \cite{Seg07} an almost $40\,$MHz frequency shift (at the liquid helium temperature), 
caused by pulses of the infrared laser ($1550\,$nm wavelength) modulated at twice the resonator 
eigenfrequency $7.74\,$GHz. The reported laser power was  $27\,$nW.

 Different PA-DCE schemes were proposed recently in the frameworks 
of the so-called {\em circuit QED\/} (which uses superconducting qubits as `artificial atoms' 
coupled to microwave resonators \cite{Sasha,Wall09,Girvin09,Sand09}). 
An idea to use quantum resonant oscillatory contours or Josephson junctions with time-dependent parameters
(capacitance, inductance, magnetic flux, critical current, etc.) to observe the DCE was put forward many years ago
\cite{Man91,DMM93}. But concrete schemes were proposed only recently.
One of them was reported in \cite{Taka08}. Its principal part is 
a double rf-SQUID system whose Josephson critical current can be controlled by 
an external time-dependent magnetic flux.
Another scheme was proposed in \cite{Johan09}. It uses the coplanar waveguide terminated by a tunable
(also by an external magnetic flux)  superconducting quantum 
interference device (SQUID), which is equivalent to a short-circuited transmission line with a
tunable length simulating a tunable mirror. It was suggested to detect the 
flux of microwave radiation going along the transmission line outwards.
The evaluations give the photon production rate $10^5$ photons per second in the $100\,$MHz
bandwidth around the central frequency $9\,$GHz. The necessary temperature should be below
$70\,$mK. 

However, both the proposals did not contain
any information concerning one crucial detail: how to change
the parameters at the time scale shorter than the period of
the EM field mode (for example, faster than 100 ps in the
case of [102])?
A possibility of fast tuning the field in the microwave resonator was demonstrated 
experimentally in \cite{Sand08}, where 
the device consisted of a quarter wavelength coplanar waveguide resonator terminated to ground via
superconducting quantum interference devices (SQUIDs) in series.
The SQUID inductance was varied by applying an external magnetic field.
It was shown that the resonance frequency of an order of $4.8\,$GHz can be changed by about $740\,$MHz
by applying the magnetic flux pulses whose duration was of the order of $10\,$ns. However, although this time is smaller than
the photon lifetime in the cavity with the quality factor $10^4$, it is still two orders of magnitude
bigger than the duration of pulses necessary for the observation of the DCE.

Many effects of the circuit QED can be understood in the frameworks of a simple model
with the interaction Hamiltonian 
$\hat{H}_{int}=\hbar g(t)(\hat{a}+\hat{a}^{\dagger})(\hat{\sigma}_+ +\hat{\sigma}_-)$
\cite{Ciuti09} [where $\hat{\sigma}_+$ and $\hat{\sigma}_-$ are the raising and lowering
operators describing atomic (electron) excitations]
 or its generalizations \cite{Sasha}. 
 Since this Hamiltonian is formally {\em linear\/} with respect to the photon creation and
 annihilation operators $\hat{a}^{\dagger}$ and $\hat{a}$, one can expect, in principle, 
 a higher photon generation rate than in the case of the frequency variation (where the
 interaction Hamiltonian is {\em quadratic\/} with respect to 
 $\hat{a}^{\dagger}$ and $\hat{a}$), provided a strong modulation of the 
effective Rabi frequency $g(t)$ can be achieved. It seems that such a possibility exists:
see the next subsection.

The photon generation due to the time-dependent
variations of properties of {\em dielectric\/} transparent  media was  considered by many authors
\cite{DKNPR,Lob91b,Hizh92,SaHy96b,Shim97,Eber99,Mend05,Carus08,BB08}. 
This kind of phenomena also belongs to the PA-DCE class.
However, realistic experimental schemes were proposed only recently. They are
described in the next subsection.

\subsection{Emission of Casimir radiation outside a cavity}
\label{sec-out}

The photon flux radiated {\em outside\/} the one-dimensional Fabry--P\'erot cavity 
with harmonically oscillating semitransparent boundaries was calculated in \cite{Lamb}.
Using the spectral approach, it was shown that
the radiation can be essentially enhanced under the resonance conditions,
comparing with the case of a single oscillating mirror.
It was suggested recently \cite{Dezael} that the motion of boundary could be
imitated by putting inside the cavity a thin nonlinear crystal (the thickness $0.1\,\mu$m,
the non-linear susceptibility $\chi^{(2)} \sim 10^{-11}\,$mV$^{-1}$),
pumped by an optical beam of frequency $f=\Omega/2\pi \approx 3\times 10^{14}\,$Hz 
($\lambda=c/f = 1\,\mu$m) and
 power about $1\,$W, focalized over an area $A=10^{-10}\,$m$^2$
(the `Casimir' photons can be distinguished from the 
pump ones due to the orthogonal polarizations).  
The evaluations were made in the framework
of the {\em one-dimensional\/} model of the cavity (because the {\em equidistant\/} spectrum of
eigenfrequencies was used explicitly). Consequently, the cavity must be very small: its length
$L$ must obey the inequality $L \ll \sqrt{A} = 10\,\mu$m.
Actually, this inequality together with the resonance conditions can be satisfied in the case 
involved only for $L=\lambda$. These values of parameters result in
the amplitude of variation of the effective cavity length $\Delta L \sim 10^{-12}\,$m 
and the relative maximum velocity of the equivalent moving boundary
$\beta =v/c \sim 10^{-6}$. Then the formulas derived in \cite{Lamb} give the following average total flux
of `Casimir' photons leaving the cavity with finesse $F=10^4$ after time $t$:
$\langle N^{out}\rangle/t = \beta^2 F\Omega /(3\pi) \sim 10^5$ photons per second
(accidentally the same number as for the microwave photons in the scheme of \cite{Johan09}).
As a matter of fact this is not a big number, because it means that photons are emitted with
intervals about $10^{-5}\,$s, so it is necessary to wait for $100\,\mu$s in order to register
about 10 photons. This is explained by the low stationary mean number of photons {\em inside\/}
the cavity: according to \cite{Lamb} $\langle N^{in}\rangle = 2(\beta F)^2  /(3\pi^2) \sim 10^{-5}$,
and this evaluation follows also from formula (\ref{Nlin}), if one identifies $t$ with the relaxation time of the
leaking cavity $t \sim F/\omega_1$ and puts $\kappa = \Delta L/(2L) = 5\times 10^{-7}$.
For comparison,
in the MIR experiment (discussed in subsection \ref{sec-MIR}) 
it is expected to generate from $10^3$ to $10^4$ {\em microwave\/} quanta
after $1000-2000$ laser pulses of the total duration $0.2-0.4\,\mu$s and the total energy about
$10-20\,$mJ. To emit the same amount of photons from the Fabry--P\'erot cavity under consideration one needs from
$10$ to $100$ mJ in the pumping laser beam.

A possibility of emission of the infra-red photons from semiconductor microcavities
with a time-modulated vacuum Rabi frequency was studied theoretically in \cite{Ciuti05,Ciuti07}.
The authors considered a planar Fabry-P\'erot
resonator embedding a sequence of many identical quantum wells 
doped with a two-dimensional electron gas.
It was shown that such a system permits one to obtain an ultrastrong light-matter coupling
(namely, the ratio of the Rabi frequency to the frequency of the intersubband transition can be
of the order of $0.1$),
which can be easily tunable by applying to the metallic mirrors a bias voltage
(which changes the density of the two-dimensional electron gas).
As was demonstrated in \cite{Gunter09}, the Rabi frequency can be changed 
on a timescale shorter than the cycle of emitted light. The cavity contained 50 identical undoped
GaAs quantum wells. The effective thickness of the structure corresponded to $\lambda/2$  for
the intersubband absorption line $\lambda=11\,\mu$m (the period $T_0=37\,$fs).
The electronic transitions from the valence band into conduction subband
were activated by near-infrared $12\,$fs control pulses with the photon energy $1.55\,$eV
and the intensity up to $0.1\,$mJ\,cm$^{-2}$.
The authors wrote in \cite{Gunter09} that the number of vacuum photons released 
per pulse could be of the order of $10^3$. However, this number seems to be exaggerated, because it strongly
depends on the modulation amplitude of the Rabi frequency. For example, Fig. 3 of \cite{Ciuti07} shows
that for an extremely big modulation amplitude $20\,$\% (which hardly can be achieved, although no evaluations
of the realistic values of this important parameter were given)
 the rate of emitted photons $dN/dt$ does not exceed
$10^{-2}\omega_{12}$, where $\omega_{12} =2\pi/T_0$. For $T_0=37\,$fs one obtains
$dN/dt \sim 2\times 10^{12}\,$phot/s or $0.07\,$phot/pulse. For the cavity quality factor $10^5$
this would result finally in about $10^3-10^4$ infra-red photons, i.e. more or less the same number 
as expected for other schemes. Therefore this scheme also seems to be perspective.

\subsection{Detection and photon statistics}

The distribution function of photons generated via the DCE is {\em non-thermal\/} 
\cite{DK96,DA99}. 
The probability $f(m)$ to generate exactly $m $ photons 
in the single mode case is given by the formula \cite{DCas60,D09} 
$f_{DCE}(m) \approx (2\pi{\cal N}m)^{-1/2}\exp[-m/(2{\cal N})]$, 
where ${\cal N}$ is the mean number of photons (this simple result holds for $m \gg 1$ and ${\cal N}\gg 1$). 
On the other hand, the thermal distribution
has the form $f_{th}(m) \approx {\cal N}^{-1}\exp[-m/{\cal N}]$ if ${\cal N}\gg 1$.
This example shows the difference between the DCE and the so-called Unruh effect
\cite{Unruh,Mend08,Matsas08}.

But how to detect the quanta of the EM field generated due to the DCE? One of possibilities 
(quite standard for the
cavity QED experiments \cite{Har}) could be to pass a beam of Rydberg atoms through the
cavity \cite{DK96,D95}. To achieve a better sensitivity, it was proposed \cite{Onof06,Onof-JPA} 
to send an ensemble of population-inverted atoms, using the effect of {\em superradiance}.
The electron beams were proposed for this purpose in \cite{Sar08}.
The simplest theoretical models of the detector are the two-level systems 
(whose interaction with the resonance field mode is described
by means of the Jaynes--Cummings model with time-dependent parameters \cite{DK96,D95,Law95,Janow03,Fedot00})
or harmonic oscillators \cite{DK96,D95}.
It was shown that the field--detector interaction can change significantly both the photon
generation rate and the photon distribution function \cite{DK96,D95}.
But this subject needs more thorough investigations using more realistic models. 
Another important problem is related to the statistics of {\em counts\/} 
by detectors (it can be quite different from the photon statistics in the field mode due to
effects of counting efficiency, dead times, etc.). This problem 
(with respect to the DCE experiments) was not considered at all until now.

\section{Conclusion}
\label{sec-fin}

This brief review of the most important results obtained by different groups 
of theoreticians and experimentalists for a few past years shows
that experimental observations of different manifestations of the dynamical Casimir effect 
are quite possible and can be expected in the nearest future.

\section*{Acknowledgments}

The Brazilian agency CNPq is acknowledged for partial
support of this work.

\end{document}